\documentclass[12pt]{article}
\pdfoutput=1
\usepackage{amssymb}
\usepackage{amsmath}
\usepackage{amsthm}
\usepackage{cancel}
\usepackage{slashed}
\usepackage{fullpage}
\usepackage{color}
\usepackage{braket}
\usepackage{graphicx}
\usepackage{multirow}
\usepackage{verbatim}
\usepackage{cite}
\usepackage{bigints}
\usepackage{simplewick}
\usepackage{authblk}

\usepackage{caption}
\usepackage{subcaption}

\usepackage{amsfonts,latexsym,amsthm,amssymb,amsmath,amscd,euscript,graphicx,fontenc,eufrak,float, url,hyperref,makecell}
\usepackage{fullpage}

\theoremstyle{definition}

\theoremstyle{remark}

\definecolor{darkgreen}{rgb}{0.0,0.5,0.0}
\usepackage{hyperref}
\hypersetup{
    linktocpage,
     colorlinks,
     citecolor=darkgreen,
     linkcolor= darkgreen,
     urlcolor=darkgreen
}

\newcommand{\sicbar}{\overline{\text{SIC}}}

\newcommand{\Q}{ {\mathcal{Q}}}

\newcommand{\Qk}{\Q_\kappa}

\definecolor{mply}{rgb}{0.75,0.75,0}
\definecolor{mplm}{rgb}{0.75,0,0.75}
\definecolor{mplc}{rgb}{0,0.75,0.75}
\definecolor{mplg}{rgb}{0,0.5,0}
\definecolor{mplr}{rgb}{1.0,0,0}
\definecolor{mplb}{rgb}{0,0,1.0}
\definecolor{mplo}{RGB}{253,164,40}

\newcommand{\nc}{\newcommand}

\nc{\on}{\operatorname}
\title{Jet Charge and Machine Learning}

\author{Katherine Fraser}
\author{Matthew D. Schwartz}
\affil{\small \emph{Department of Physics, Harvard University, Cambridge, MA 02138, USA}}

 \date{}
\begin{document} 

\maketitle

\begin{abstract}
Modern machine learning techniques, such as convolutional, recurrent and recursive neural networks, have shown promise
for jet substructure at the Large Hadron Collider. For example, they have demonstrated effectiveness at
 boosted top or $W$ boson identification or for quark/gluon
discrimination. We explore these methods for the purpose of classifying jets according to their electric charge. We find that both neural networks that incorporate distance within the jet as an input and boosted decision trees including radial distance information can provide significant improvement in jet charge extraction over current methods. Specifically, convolutional, recurrent, and recursive networks can provide the largest improvement over traditional methods, in part by effectively utilizing distance within the jet or clustering history. The advantages of using a fixed-size input representation (as with the CNN) or a small input representation (as with the RNN) suggest that both convolutional and recurrent networks will be essential to the future of modern machine learning at colliders.
\end{abstract}

\newpage

\section{Introduction}
As the Large Hadron Collider, currently running at CERN, increases in luminosity, it becomes sensitive to signals of beyond-the-standard model physics
with ever smaller cross sections. These signals, particularly if they involve hadronic final states known as jets, are often buried in enormous backgrounds, so any tools that help reduce those backgrounds will be invaluable. In addition, increased clarity on jet properties and substructure will
constrain and test the Standard Model.
Over the last decade or so there has been tremendous progress in understanding jets and measuring their properties, from finding boosted top quark or $W$-jets~\cite{Kaplan:2008ie,Almeida:2008yp,Cui:2010km,Thaler:2011gf}, to looking at jet substructure~\cite{Salam:2009jx,Altheimer:2012mn}. Recently, new methods from computer science involving modern machine learning are starting to be adapted to jet physics, with remarkable early progress~\cite{deOliveira:2015xxd,Komiske:2016rsd,Komiske:2017ubm,Luo:2017ncs,Cheng:2017rdo,Metodiev:2017vrx,Butter:2017cot,Cohen:2017exh,Pearkes:2017hku,Kasieczka:2017nvn,Louppe:2017ipp,Larkoski:2017jix,Komiske:2018oaa,Macaluso:2018tck}.

In this paper, we consider how modern machine learning might help in measuring the electric charge of a jet. Doing so accurately would allow us to differentiate 
 up-quark initiated jets ($Q=\frac{2}{3}$) from anti-up-quark ($Q=-\frac{2}{3}$), down-quark  ($Q=-\frac{1}{3}$), anti-down quark ($Q=\frac{1}{3}$)
 and gluon-initiated jets ($Q=0$). This is clearly an ambitious goal, but there is already evidence that relatively simple observables, such as the $p_T$-weighted jet charge 
  \begin{equation} \label{Omdef}
    \Qk = \frac{1}{(p_T^{\rm jet})^\kappa} \sum_{j \in \mathrm{jet}} Q_j (p_T^j)^\kappa
\end{equation}
 can help. This observable, adapted from early work of Feynman and Field~\cite{Field:1977fa}, was shown in~\cite{Krohn:2012fg,Waalewijn:2012sv} 
 to have appealing  theoretical properties, such as a calculable scale-dependence. Measurements of $\Qk$ by both ATLAS and 
 CMS~\cite{TheATLAScollaboration:2013sia, Nachman:2014qma,TheATLAScollaboration:2015bgc, Aad:2015cua,CMS:2016yuu, Sirunyan:2017tyr,Tokar:2017syr}
  confirmed its utility and demonstrated that, on average, positive and negative electrically charged jets can be distinguished. Moreover, the scale-dependence predicted in~\cite{Krohn:2012fg,Waalewijn:2012sv} was confirmed experimentally~\cite{TheATLAScollaboration:2013sia}. Thus, considering that jet charge can 
  already be measured to some extent, it is natural to ask if we can do better using deep learning or other modern machine-learning ideas.

The challenge of extracting the jet electric charge is not unlike the challenge of extracting the jet color charge, namely whether a jet is quark- or gluon initiated. Quark/gluon jet discrimination also has a long history~\cite{Pumplin:1991kc, Lonnblad:1990bi, Acton:1993jm, Alexander:1991ce}.  Some Monte-Carlo studies
showed good potential for the LHC~\cite{Gallicchio:2012ez,Gallicchio:2011xq}, and experimental studies showed feasibility while also uncovering some challenges (such as the untrustworthiness of the simulations particularly for gluon jets, though some studies have avoided this issue by developing methods to  train the network directly on data)~\cite{Aad:2014gea,Komiske:2018oaa}. One of the first modern-machine-learning jet physics papers~\cite{Komiske:2016rsd}  showed, using convolutional neural networks (CNNs) and jet images~\cite{Cogan:2014oua,Almeida:2015jua}, a significant improvement over previous quark-gluon discrimination benchmarks (see also~\cite{ Almeida:2015jua,Baldi:2016fql,Guest:2016iqz,Kasieczka:2017nvn}).  Work on testing this method in experiment is ongoing~\cite{ATL-PHYS-PUB-2017-017}.
  
While the jet images approach is powerful, it involves embedding the jet data in a very high dimensional representation. For example, a jet may have 50 particles, so it is characterized by 50 three-momenta, or 150 degrees of freedom. A $33 \times 33$ jet image has 1089 degrees of freedom. Alternatives to jet images are methods such as recursive and recurrent neural networks. Thus besides developing a powerful jet charge discriminator, one goal of this paper is to compare the performance of different network architectures on jet charge extraction and quark/gluon discrimination.

Recurrent neural networks have been considered for collider physics applications in~\cite{Louppe:2017ipp,Butter:2017cot}. In particular~\cite{Butter:2017cot} 
considered the application of a particular recurrent framework for top-tagging and found comparable performance to a jet-images based approach~\cite{Kasieczka:2017nvn}. A challenge with recurrent networks is how to sort and process the inputs. One option is to use 4-vectors, as in~\cite{Louppe:2017ipp,Butter:2017cot}. In~\cite{Butter:2017cot} the 4-vectors were processed with a network constructed to respect their Lorentz structure.
We will instead consider recurrent network inputs containing various distillations of the 4-vector input, such as into the energy of the jet, or the clustering distance to the jet axis. 

The paper is divided into two parts: a discussion of the networks in Section~\ref{sec:nets} and a discussion of the results in Section~\ref{sec:results}. A summary and broader conclusions are in Section~\ref{sec:conc}. 

\section{Methods \label{sec:nets}}
For this study, we simulated quark and gluon jets using {\sc pythia} 8.226\cite{pythia} with the default tune. Although simulations may not be completely trustworthy, the relative efficacy of different methods can still be tested using Monte-Carlo. 
For concreteness, we focused on discriminating up-quark-initiated jets from down-quark-initiated jets, though in principle we'd expect similar results for anti-down versus anti-up discrimination. These jets were selected as the hardest jet in $uu \to u u$
or $dd\to dd$  dijet events in $p p$ collisions with $\sqrt{s} = 13$ TeV.
For quark/gluon discrimination, the processes $pp\to qq$ and $pp \to gg$ were used and again the hardest jet taken.
Jets  were clustered with the anti-$k_T$ algorithm with $R=0.4$, and only jets with pT between 100-120 GeV and 1000-1200 GeV were selected. Final state particles with  $|\eta| > 2.5$ and neutrinos were discarded. 100,000 of each type of event were generated and 80\% were used for training, 10\% were used for validation, and 10\% were used for testing.

We consider a number of different machine learning methods and compare them to jet charge. 

\subsection{Convolutional networks (jet images)}
From each event we constructed a jet image, following the procedure of \cite{Komiske:2016rsd}. We considered two-channel jet images, where each channel encodes different input information. Each channel of the image is constructed by putting a $\Delta \phi \times \Delta \eta = 33\times 33$ pixel box around each jet. For the first channel, the pixel intensity is given by the sum of the transverse momenta of all particles entering that pixel.
For the second channel, the pixel intensity is given by the $p_T$-weighted jet charge, as in Eq. \eqref{Omdef}, for a given $\kappa$.
 During image generation, the image is centered and the momentum channel is normalized by the sum of the momenta of all the particles in the jet. The same preprocessing and data augmentation as \cite{Komiske:2016rsd} was used on the images. This preprocessing includes zero centering and dividing by the standard deviation, and data augmentation includes translations by one pixel in each direction and reflections. A random rotation was tested but did not improve performance. 

The images are processed with a convolutional neural network, as in~\cite{Komiske:2016rsd}. Our basic CNN  consisted of three layers of convolutional filters, one dense layer with 64 neurons, and a final dense layer with 2 neurons. Each convolutional layer is followed by a maxpooling layer and a dropout layer and the first dense layer is followed by a dropout layer. The dropout was 0.18 for the first layer and 0.35 for the other layers. The convolutional layers and the first dense layer have ReLU activations, while the second dense layer has a softmax activation. The network was trained in batches of 512 for 35 epochs with an early stopping patience of 5 epochs, using the Adam algorithm and categorical crossentropy loss function. Each layer had 64 filters. The filter size was 8 $\times$ 8 pixels for the first layer and 4 $\times$ 4 pixels for the other layers. 

Other network parameters were also tested. For two-channel images, we considered the effect of modifying the step size and decay within optimization, batch size, the dropout after each layer, filter size, number of filters, size of the maxpooling layer, activation function for the convolutional layers (selu), early stopping patience, and optimizer (SGD, RMS Prop, Adagrad). We also experimented with modifying network structure by adding additional convolutional layers at the beginning of the network and extra dense layers after the convolutional layers. The configuration detailed above was the most effective. 

In addition to modifying network structure, we tried modifying the content of the channels of the network by adding a third channel with more information. Adding a third channel with the number of neutral particles did not improve results. Adding a third channel with jet charge per pixel for a second $\kappa$ value did improve training speed, but not results (see Fig.~\ref{fig:channels3}). Furthermore, with a second $\kappa$ value the dropout value needed to be higher to avoid over training. We also tested the results with only a single input channel (also displayed in Fig.~\ref{fig:channels3} for a single jet charge channel). We also tested the network with only a pT-input channel, but this network was unable to distinguish the up quark initiated versus down quark initiated jets. 

We also tested another CNN configuration (known as a residual CNN), modeled on~\cite{He:2015}, which won the ILSVRC 2015 image recognition challenge. Although the residual CNN uses the same physical inputs as our basic CNN, in other applications residual CNNs have been shown to train faster and more consistently than more basic CNNs on the same data set. What distinguishes the residual CNN from our basic CNN is that it uses shortcut connections that connect a given layer to some previous layers while skipping one or more intermediate layers. We use the identity mapping as our shortcut connection, so that the output of each convolutional layer except the first is added to the input of that layer before it is passed to the next layer, which in~\cite{He:2015} was shown to improve classification in previous image recognition challenges. Following the observation in~\cite{He:2015} that residual CNNs show more improvement for deeper networks, we use a deeper network than our other CNN. We use seven layers each with 64 filters of size $2 \times 2$. We use smaller filters than our other CNN because of memory constraints for the deeper network. There is a maxpooling layer of size 4 after the fourth and eighth layers, and a maxpooling layer with size two after the seventh layer. As with the shallower CNN, the convolutional layers are immediately followed by two dense layers, the first with 64 nodes and the second with 2 nodes. Dropout of 0.2 was used after each maxpooling layer, and dropout of 0.1 was used after the first dense layer. These parameters were determined by a scan of selection of parameters. Other hyperparameters are the same as in the shallower network.

\subsection{Recurrent  networks}
We also tested a recurrent network (RNN) with various different inputs. 
In an RNN, each layer consists of multiple nodes with a set of hidden weights. Both the input and output of each layer is an ordered sequence of vectors, where each vector in the sequence has fixed length but the length of the sequence itself is arbitrary. In particular, for the input layer of our RNN, each vector corresponds to a single particle in the jet, and the sequence of vectors corresponds to the list of particles in the jet. 
Network performance is sensitive to the order of the input vectors. 

We implemented a recurrent network using keras \cite{chollet2015keras} with a Theano backend. It consists of 11 gated recurrent unit layers (GRUs), followed by a dense layer with 64 nodes and a dense layer with two output nodes. The number of nodes in each GRU layer decreases from 100 to 5, where the number of nodes in each of the first ten layers decreases by ten from the previous layer. Each GRU layer except the last returns a sequence of vectors, and the last returns the average of the sequence of vectors. The number and size of the GRU layers were determined by trail and error. An additional dense layer of 64 units was tested but decreased classification effectiveness.
Long short-term memory (LSTM) layers were also tested and performed similarly to GRU layers.
Additionally, we tested various different input representations. We considered combinations of azimuthal angle $\phi$, pseudorapidity $\eta$, $p_T$, charge $Q$ and various distance measures, which is discussed more thoroughly in the results section.

A batch size of 6000 was used for training with step size of $0.005$. Other batch sizes were tested. We found that for small batch sizes training was very slow and non-convergent (batch sizes less than about 4000 are unable to distinguish the two samples). Training improved with larger batch size up to 6000. A step size of 0.001 was also tested but training was more consistent with a smaller step size. Optimization was performed using the Adam algorithm with a categorical cross entropy loss function and early stopping patience of 3 and a maximum of 100 epochs. In order to use keras a maximum sequence length must be set for the input layer. We set this to 40 particles for 100 GeV and 120 particles for 1000 GeV, so that it would include enough particles not to affect training.

Additionally, we tested another configuration which is discussed below for completeness. This modification to the RNN had a last dense layer with a single output node attempting to predict charge itself (instead of classification). Here we used mean squared error as the loss function (as categorical cross entropy only makes sense for classification) and a linear activation function (instead of a ReLU) for the second of the two dense layers (because we wanted to be able to predict negative values). This network performed so similarly to the classification case that we do not discuss it further in the results section. 

\subsection{Recursive network}

A recursive network (RecNN) is similar a recurrent network (RNN), with the key difference that the order of the inputs is different in the two cases. In a recurrent network, the vectors for each input particle in a jet are ordered in a sequence (for example, the particles in the jet might be ordered by decreasing pT or increasing distance from the jet axis). In particular, each computation depends directly only on the input vector (the particle itself) to that step and the internal hidden state after the previous particle in the jet. In contrast, recursive networks can have more complicated dependency structures. Rather than applying the same set of weights to every vector in a sequence, particles are fed to the recursive network in an order given by a more complicated data structure, such as a tree (in our network, this tree is determined by clustering history). 

The architecture of our recursive network is modeled after  \cite{Louppe:2017ipp}. A recursive embedding,  given by Eqs. (2) through (4) of  \cite{Louppe:2017ipp}, with $v_{i(k)}$ consisting of $p_T$,$\phi$, $\eta$ and charge $Q$, is fed into a classifier consisting of a dense layer with 64 nodes followed by a dense layer of 2 nodes. The recursive embedding consists of a single vector given by the embedding at the root node. Clustering is performed prior to passing the information to the network following the C/A, anti-$k_T$, and $k_T$ algorithms (in all cases, our jets are the same collection of particles identified with anti-$k_T$). For the input to the leaf nodes the charge $Q$ is the charge of the particle corresponding to the leaf. For the input to the interior nodes, we find the best performance when the charge $Q$ is taken to be the $p_T$-weighted jet charges of the left and right children with $\kappa = 0.2$ at 100 GeV and $\kappa = 0.1$ at 1000 GeV. A batch size of 500 was used for training; larger batch sizes increased performance and this was the maximum possible with given memory constraints on the GPU.

We also tested a simpler recursive structure inspired by jet charge, referred to as the trainable $\kappa$ NN throughout this paper. The first half of this network is recursively computed jet charge with trainable $\kappa$ values, while the second half is a dense layer with two nodes. Which $\kappa$ to use to compute the value at each node is determined by the distance from the root node in the clustering tree. For the plots in this paper, we used five $\kappa$ values, and the recursively computed jet charge of all nodes with distance greater than or equal to five from the root node were computed using the last $\kappa$. The other hyperparameters for this network were similar to those for the other recursive network.

\subsection{Other Classifiers \label{sec:BDT}}
In order to understand the improved performance of our machine learning methods, we implemented several boosted decision trees (BDTs) for comparison. We also implemented two dense neural networks (DNNs). The input to our BDTs are observables similar to jet electric charge, but also weighted by radial distance to the jet axis, which are of the form 
\begin{equation}
\Q_{\kappa, \lambda} = \frac{1}{(p_T^{\rm jet})^\kappa} \sum_{j \in \mathrm{jet}} Q_j (p_T^j)^\kappa(\Delta R_j)^\lambda
\label{eqn:QLK}
\end{equation}
We use these observables to construct three different BDTs. The first, with $\lambda = 0$, just includes 8 different values of jet electric charge where $\kappa$ runs from 0 to 0.35 in increments of 0.05. The second BDT takes $\kappa = 0$ and weights charge only by radial distance, with $\lambda$ from 0 to 0.5 in increments of 0.1. The third BDT varies  both $\kappa$ and $\lambda$ over the ranges given above, including a total of 40 observables. Our implementation of BDTs is with scikitlearn using AdaBoostClassifier with 500 samples per leaf minimum, 10 estimators, and learning rate 0.1 (based on a scan of a selection of parameters). 

We also implemented a DNN taking the 40 $Q_{\kappa, \lambda}$ observables implemented in the previous section as inputs. This network, also implemented with keras, had 5 layers each with 100 nodes and ReLU activations. The final layer of the network has two nodes with softmax activations. It was trained for 35 epochs with an early stopping patience of 5 epochs, batch sizes of 1000, and with the Adam algorithm with a step size of 0.005 (parameters were again selected based on a scan of the selection of parameters). 
In addition, we tested another DNN with a variety of filter configurations and similar parameters to the RNN in the previous section that used the p$_T$, $\eta$, $\phi$ and $Q$ of the hardest $N$ particles as input, with $N$ ranging from 5 to 10,
with 8 particles appearing optimal. This did not even perform as well as $p_T$-weighted jet charge alone, so we omit it from the results section. 

\section{Results \label{sec:results}}
In this section we present our results. The figures below include displays of the standard Receiver-Operator Characteristic (ROC) curve of the down-quark (signal) efficiency $\epsilon_s$ versus up-quark (background) efficiency $\epsilon_s$ and of the Significance Improvement Characteristic (SIC) curve of $\epsilon_s/\sqrt{\epsilon_b}$~\cite{Gallicchio:2010dq}. The SIC curves indicate approximately the improvement on discrimination significance and their peak values, $\sicbar$ gives an objective uniform quantitative measure of performance. ROC curves and SIC curves convey the same information. The beginning of the results section discusses jets with pT between 100-120 GeV, and the energy dependence section studies jets with 1000-1200 GeV pT.

\subsection{$p_T$-weighted jet charge}
\begin{figure}[t]
\centering
\includegraphics[width=80mm]{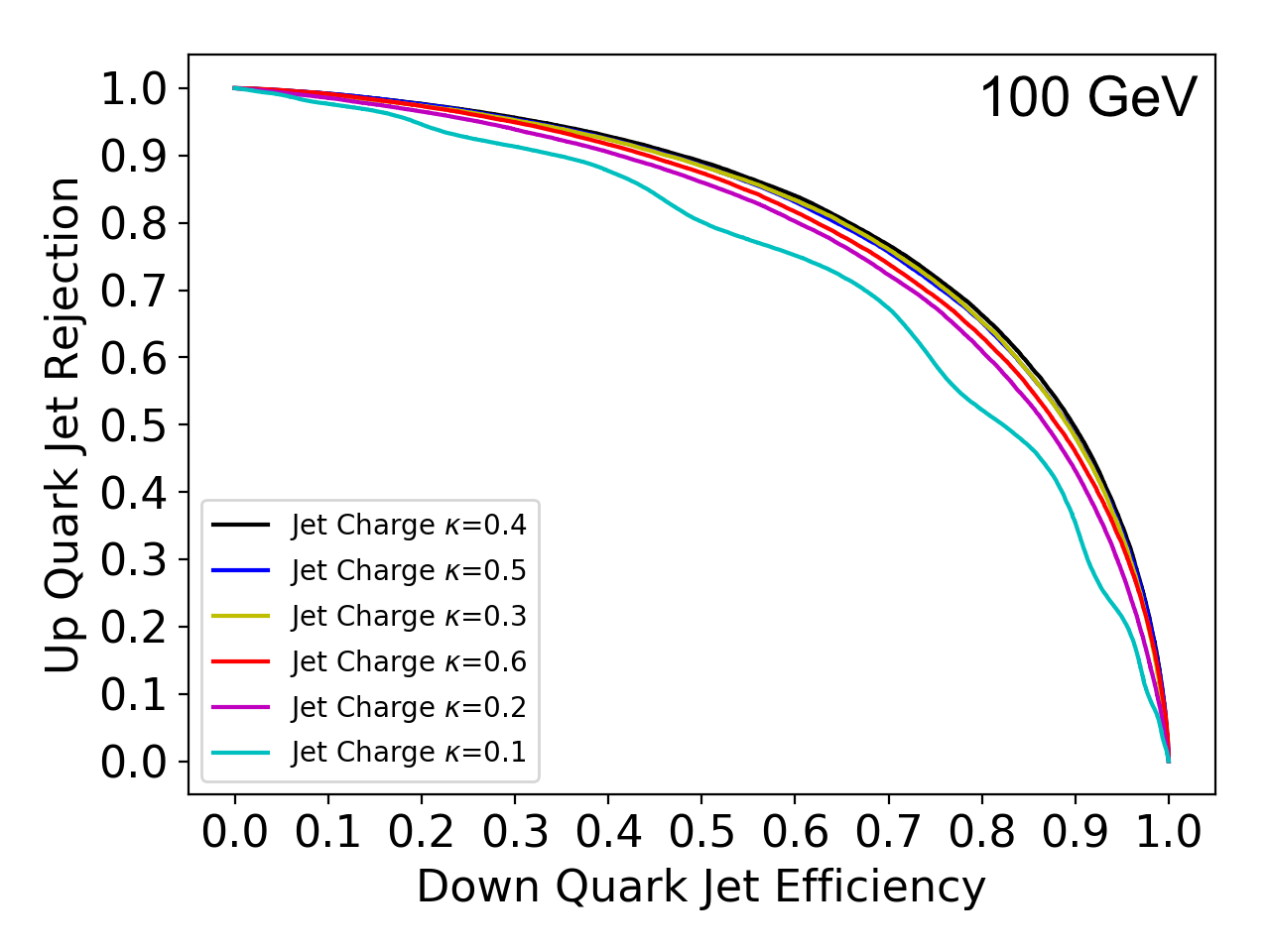}
\includegraphics[width=80mm]{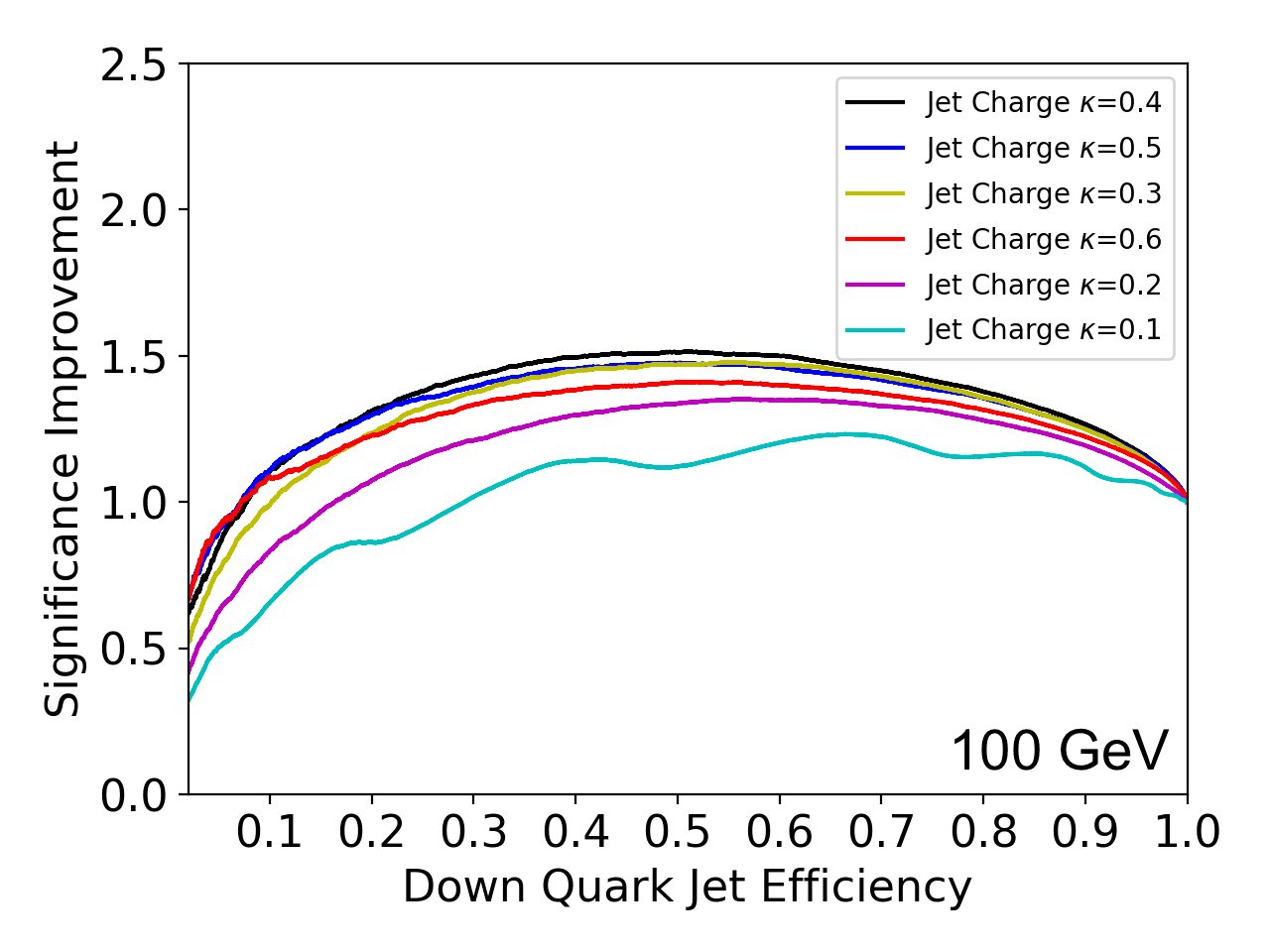}
\caption{ROC and SIC curves for $p_T$-weighted jet charge for various $\kappa$.}
\label{fig:kappas}
\end{figure}

We first evaluate the effectiveness of the $p_T$-weighted jet charge in Eq.~\eqref{Omdef} for various values of $\kappa$. The result is shown in Fig.~\ref{fig:kappas}. These results are consistent with those in~\cite{Krohn:2012fg}, showing optimal performance at $\kappa = 0.4$ with $\sicbar =1.5$.

\subsection{Jet Images}
\begin{figure}[t]
\centering
\includegraphics[width=80mm]{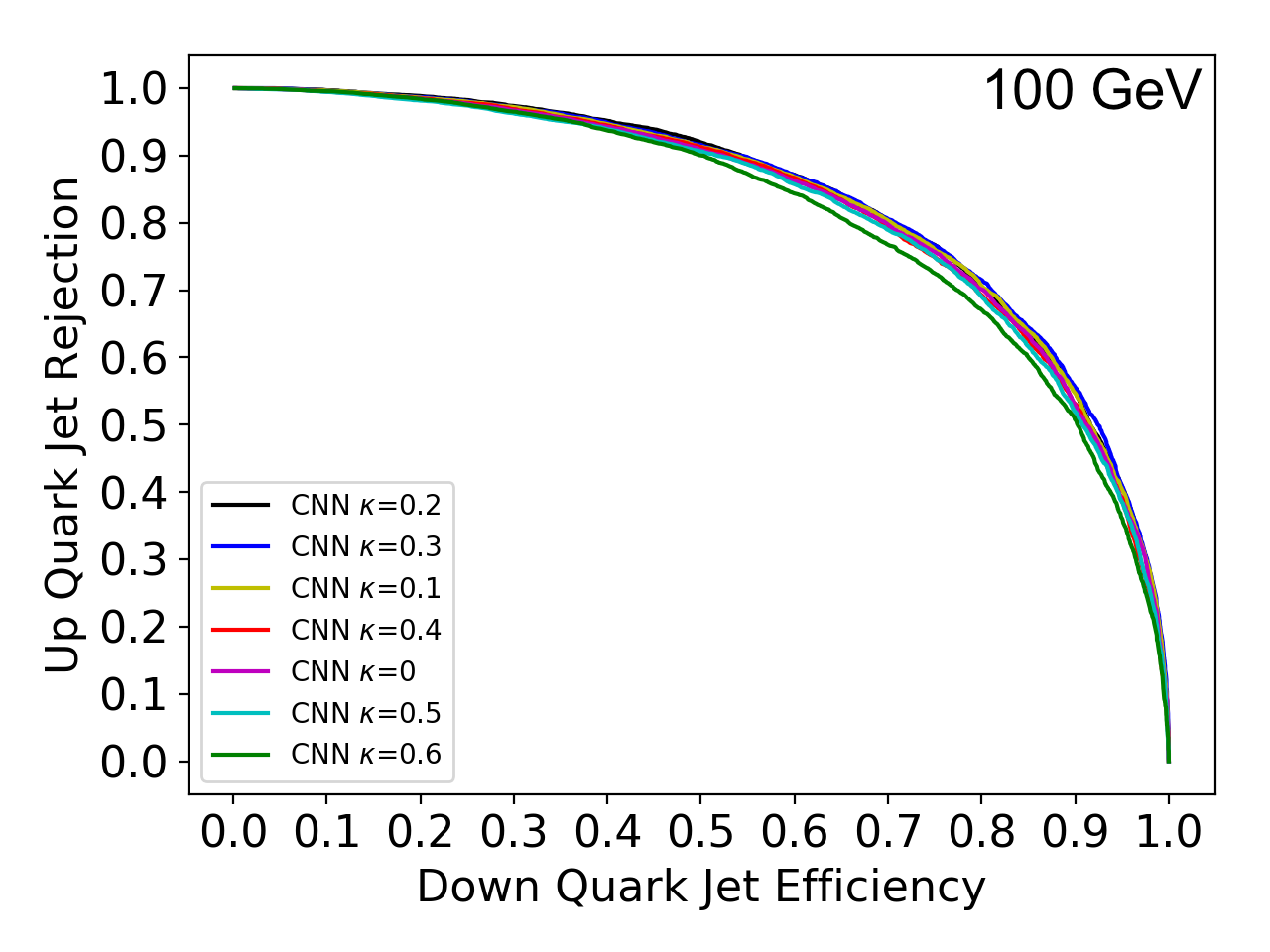}  
\includegraphics[width=80mm]{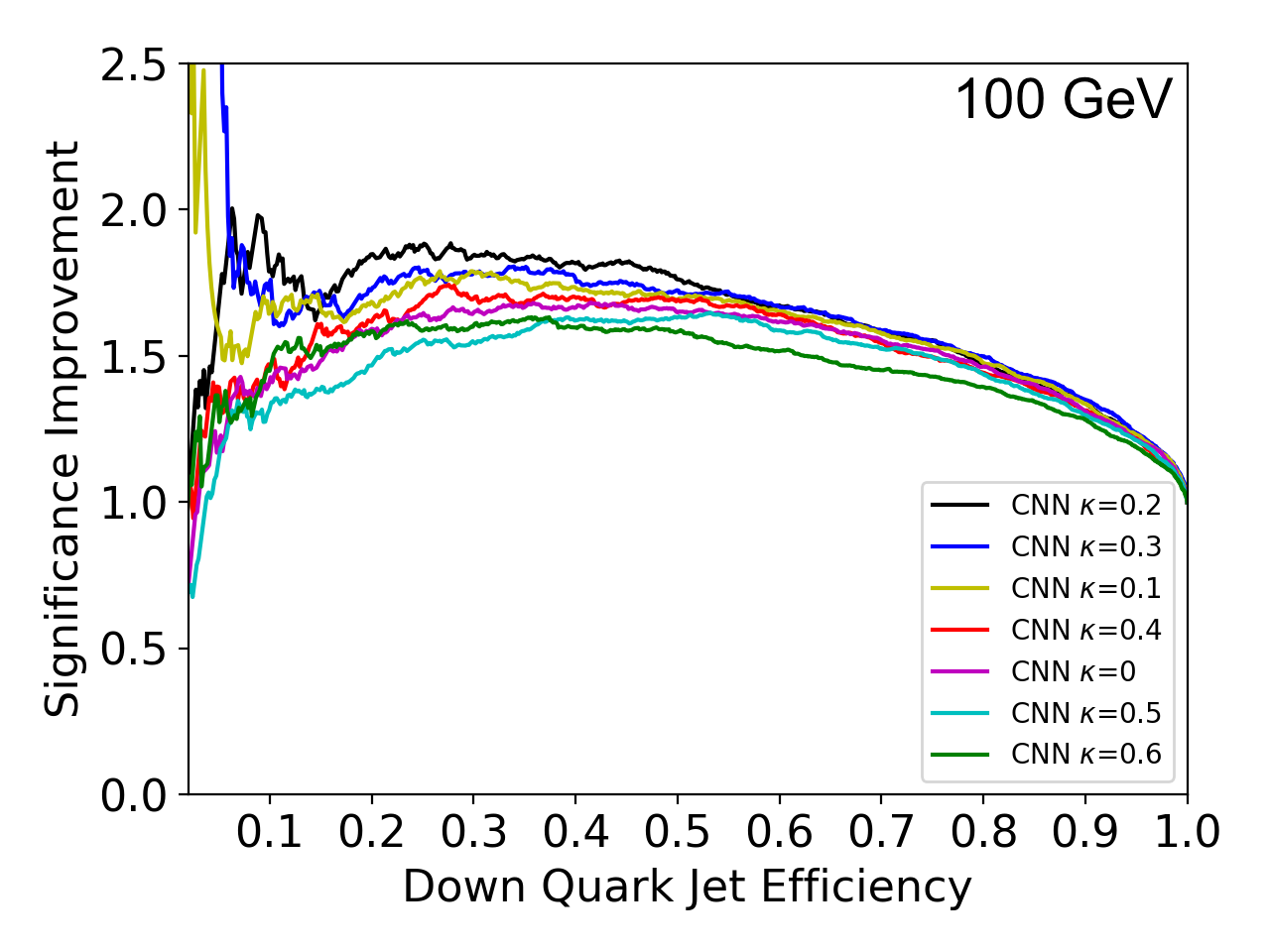}
\caption[]%
{{\small ROC and SIC curves for  jet-image based CNNs using two input images: the total $p_T$ and the $p_T$-weighted jet charge, for various $\kappa$ as listed.}}  
    \label{fig:CNN}  
\end{figure}

Next, we look at the performance of our CNN using 2-channel jet images on the same samples. The results are shown in Fig.~\ref{fig:CNN} for various $\kappa$ values. We see that the optimal $\kappa$ value for jet images is $\kappa = 0.2$, which is lower than for $p_T$-weighted jet charge. The performance of the CNN is also better with $\sicbar = 1.8$, a notable improvement. 

\begin{figure*}
    \begin{subfigure}[b]{0.5\textwidth}
        \centering
        \includegraphics[width=\textwidth]{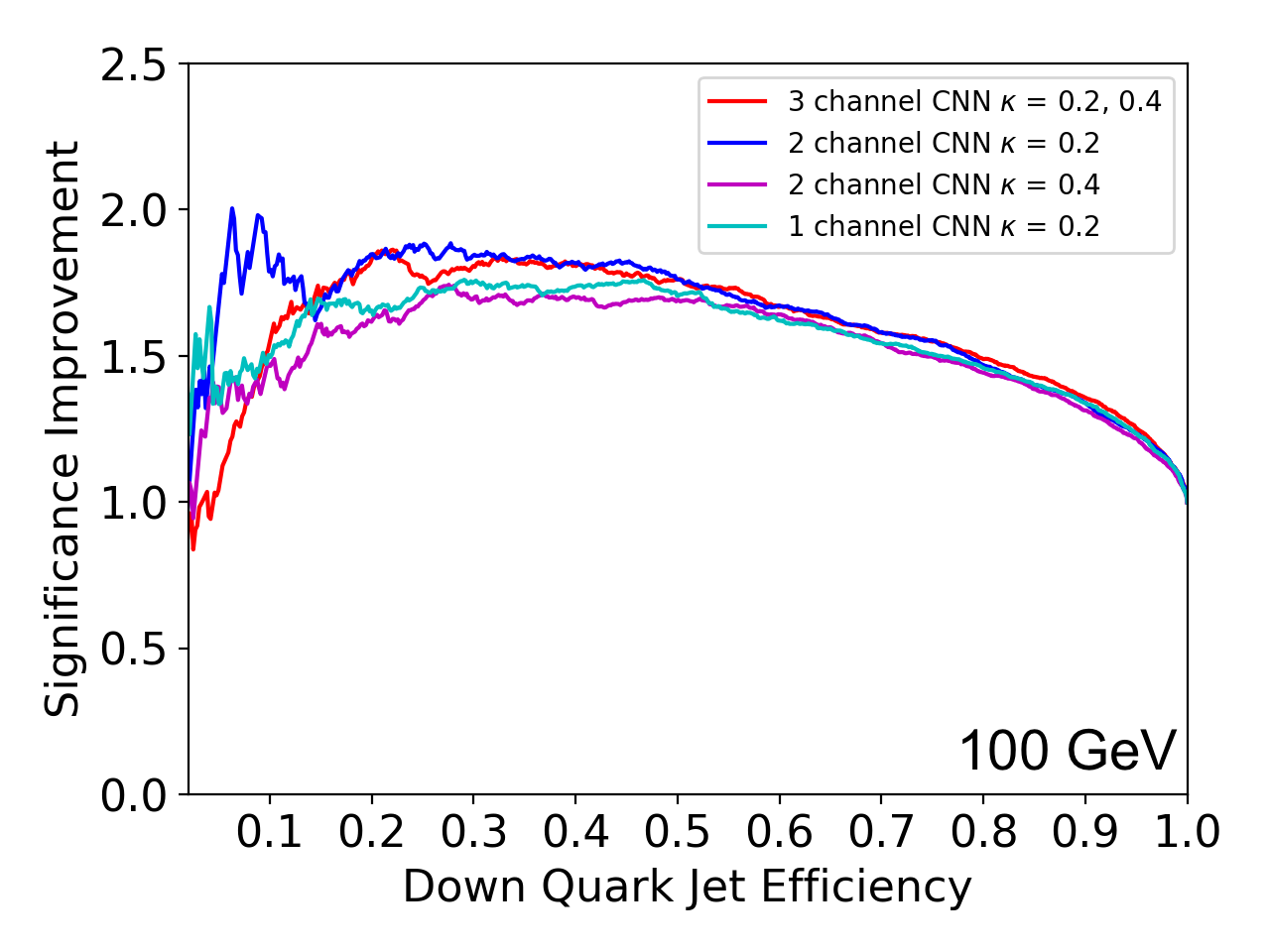}
        \caption{}  
        \label{fig:channels3}
        \end{subfigure}
        \hfill
    \begin{subfigure}[b]{0.49\textwidth}  
        \centering 
        \includegraphics[width=\textwidth]{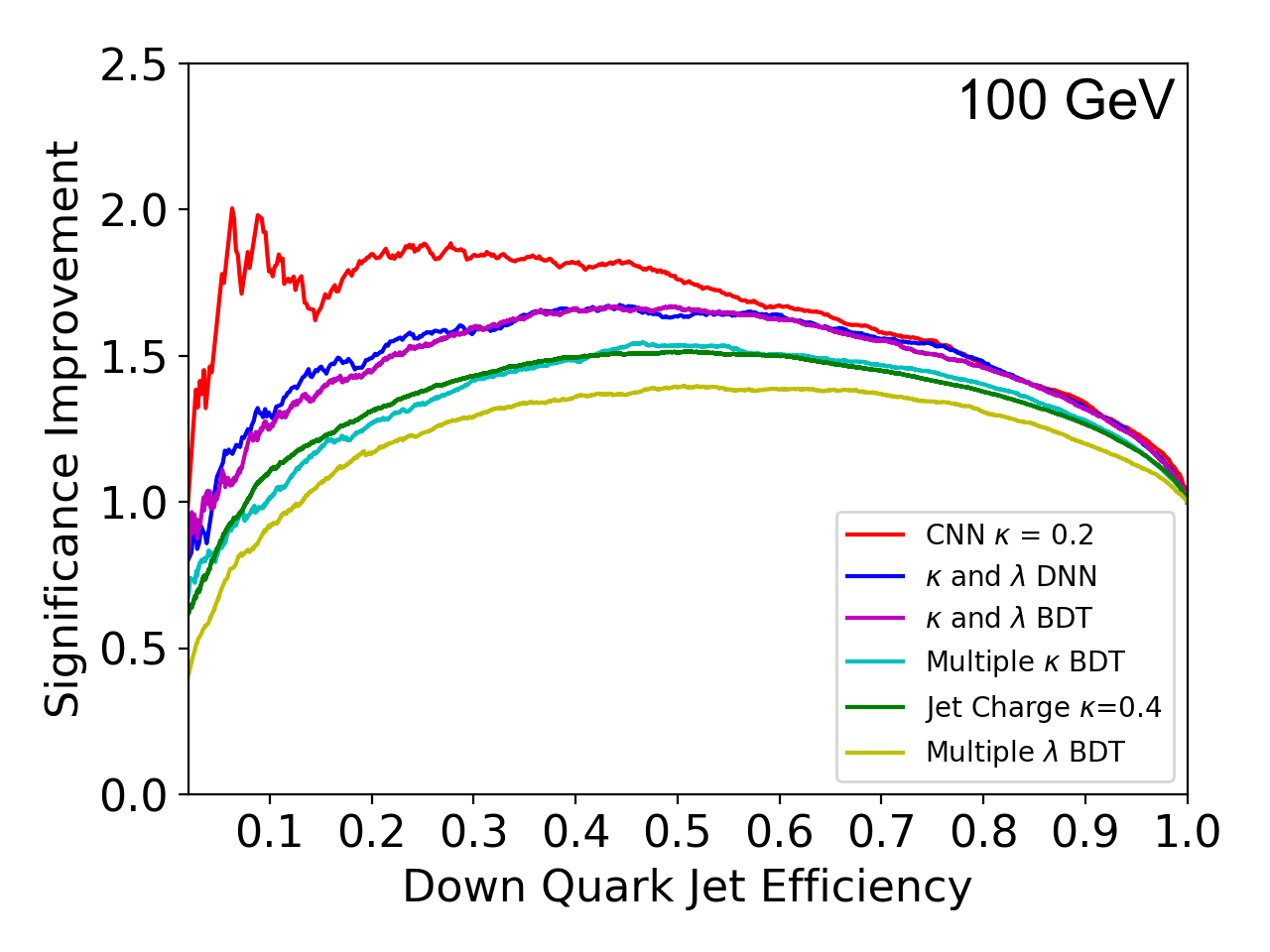}
        \caption{}
        \label{fig:CNNvscharge}
    \end{subfigure}
    \caption[]{{\small (a) Comparison of 1, 2 and 3-input channel  CNNs. Two $\kappa$ values are used for the 3 channel images, in addition to the total momentum input channel. (b) Comparison of $p_T$-weighted jet charge, CNN using two-channel jet images, several BDTs of multiple $Q_{\kappa, \lambda}$ (including cases with $\kappa = 0$ or $\lambda = 0$), and a DNN with $Q_{\kappa, \lambda}$ observables as inputs.  }}
\end{figure*}





Fig.~\ref{fig:channels3} compares the performance of the CNN with 1-channel images (no pT channel), 2-channel images (one value of $\kappa$), and 3-channel images (with the 3rd channel being the $p_T$-weighted jet change with a different value of $\kappa$). We see that adding the third layer does not improve performance. We also see that the images with a single jet charge channel are able to improve upon the observable jet charge, but do not quite match the performance of the two channel network.

Fig.~\ref{fig:CNNvscharge} compares the traditional $p_T$-weighted jet charge with $\kappa = 0.4$ to the two-channel CNN with $\kappa = 0.2$. The three BDTs of $Q_{\kappa, \lambda}$ described in section \ref{sec:BDT} are also included in this figure. The multiple $\kappa$ BDT takes jet charges as inputs ($\lambda = 0$) with $\kappa$ from 0 to 0.35 in intervals in 0.05. 
The multiple $\lambda$ BDT takes $Q_{\kappa, \lambda}$ with $\kappa = 0$ and $\lambda$ from 0 to 0.4 in increments of 0.1 as inputs. The $\kappa$ and $\lambda$ BDT and DNN also take $Q_{\kappa, \lambda}$ as input, with both $\kappa$ and $\lambda$ varied over the same intervals as described above, for a total of 40 observables.

We see that the CNN outperforms both the single $\kappa$ observable and the multiple $\kappa$ or multiple $\lambda$ value BDT. The BDT and DNN ranging over both $\kappa$ and $\lambda$ performs similarly to the CNN at high signal efficiency but does not display the same improvement at lower signal efficiency.

\subsection{Recurrent Network Results}
Next we explore the performance of a recurrent neural network with a variety of different input vectors associated to each input momentum.  
We considered
combinations of azimuthal angle $\phi$, pseudorapidity $\eta$, $p_T$, charge $Q$ and various distance measures. 
The configurations we tried were
\begin{enumerate}
\item  ($\eta,\phi,p_T,Q$) \label{it1}
\item ($\eta,\phi,p_T,Q, d_1, \dots, d_n$) where the $d_i$ are the distances to the hardest $N$ anti-$k_T$ subjets using $C/A$, $k_T$, or anti-$k_T$ distance measures \label{it2}
\item ($\eta,\phi,p_T,Q, d$)  where $d$ is the clustering-tree distance to root node \label{it3}
\item ($\eta,\phi,p_T,Q, d$)  where $d$ is the distance to the jet axis using $C/A$, $k_T$, or anti-$k_T$ distance measures \label{it4} 
\item ($p_x$, $p_y$, $p_z$, $E$, $Q$)\label{it5}
\end{enumerate}

\begin{figure*}[t]
\centering
\begin{subfigure}[t]{0.5\textwidth}
\centering
\includegraphics[width=\textwidth]{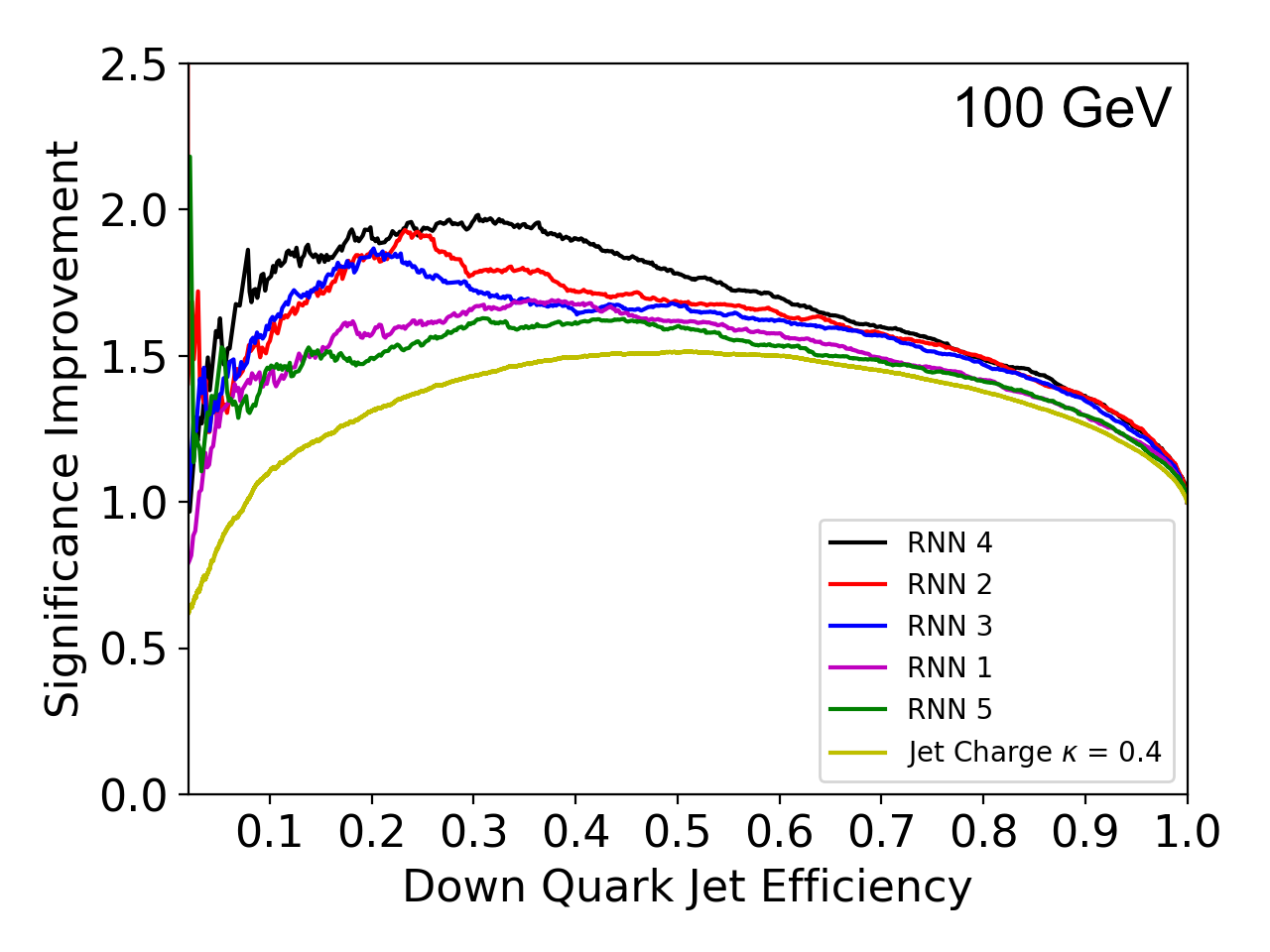}
\caption{}
\label{fig:RNNs}
\end{subfigure} 
\hfill
\begin{subfigure}[t]{0.49\textwidth}
\centering
\includegraphics[width=\textwidth]{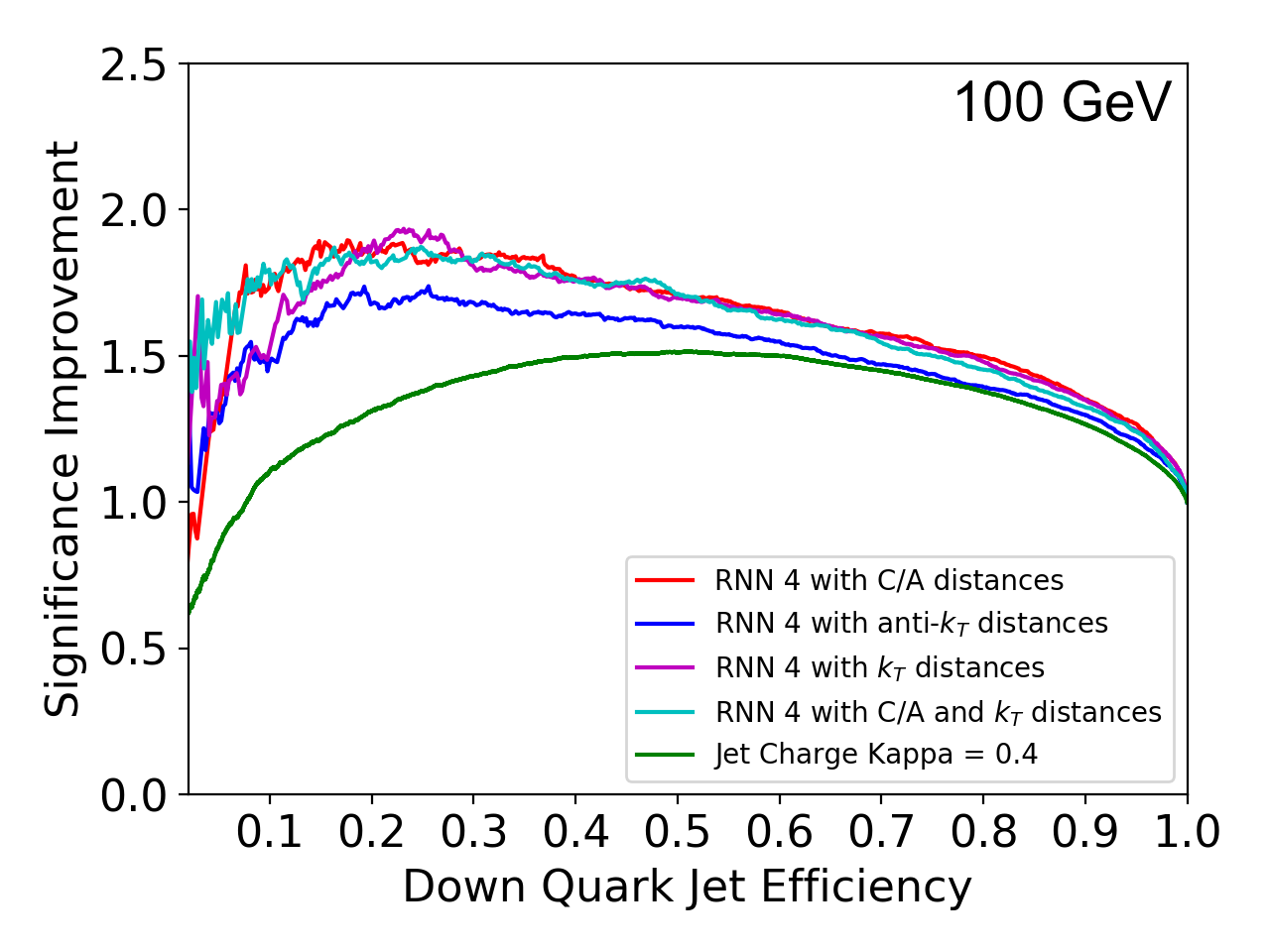}
\caption{}
\label{fig:RNN5}
\end{subfigure} 
\caption{(a) Comparison of different RNN inputs to jet charge. Configuration 2 uses N = 1. Configurations which include distance to the jet axis or hardest subjets perform better than those that do not. (b) Performance of recurrent neural networks in configuration \ref{it4} using $p_T$, charge $Q$ and distance to jet axis as inputs.}
\end{figure*}

\begin{figure}[t]
\centering
\includegraphics[width=80mm]{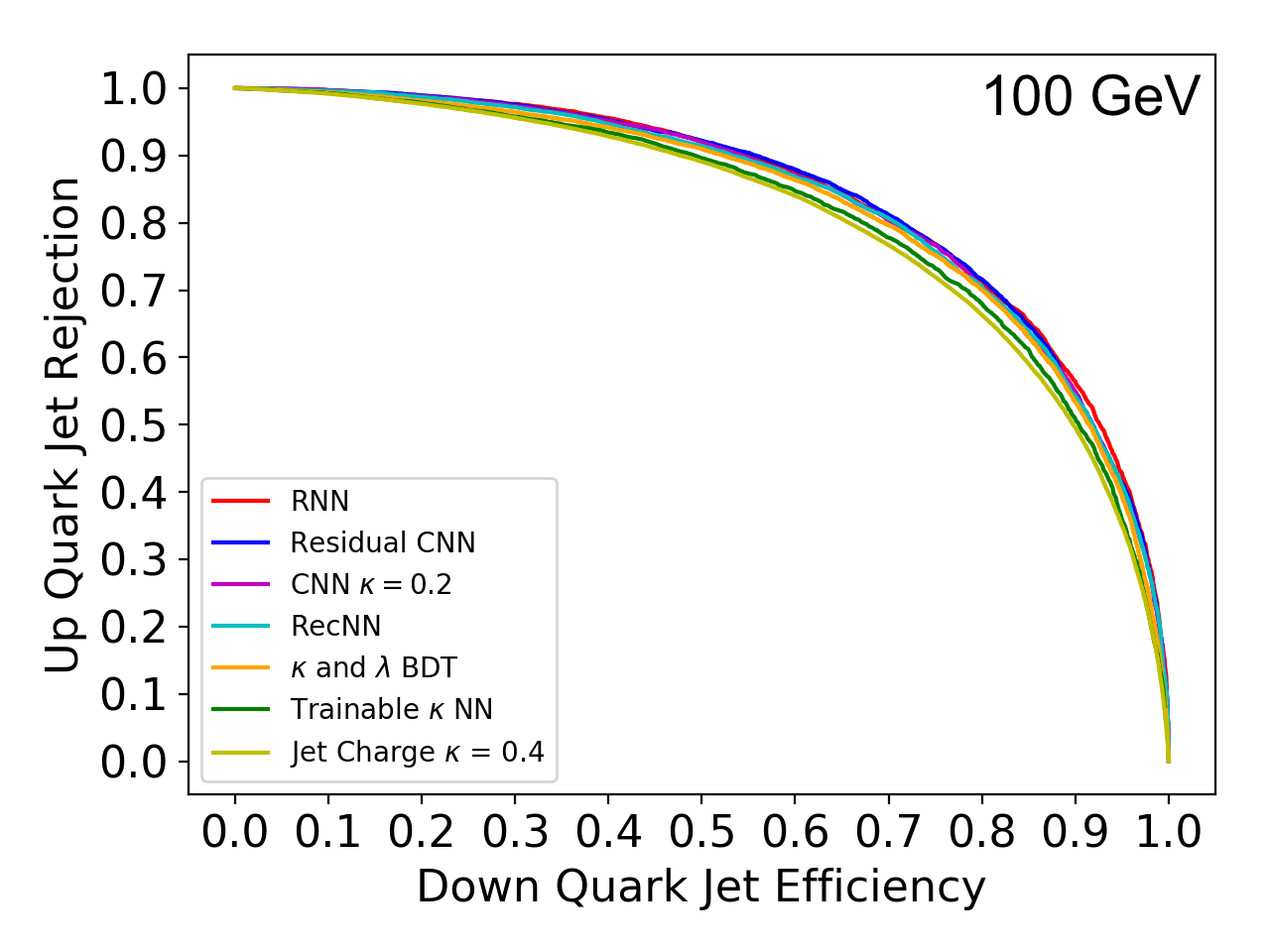}
\includegraphics[width=80mm]{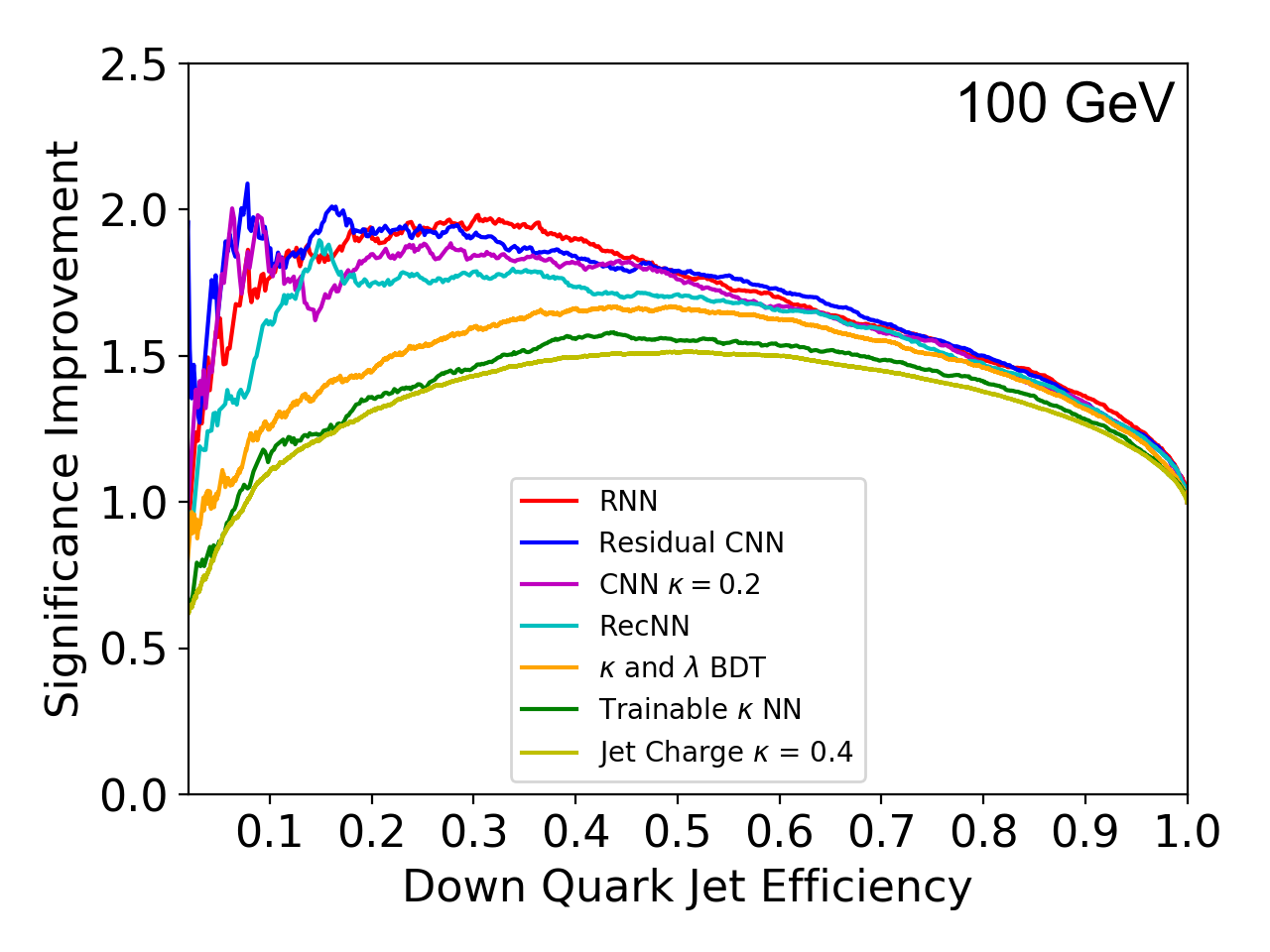}
\caption{Comparison of the $p_T$-weighted jet charge to the best performing recurrent (RNN), recursive (RecNN), and convolutional (CNN) neural networks 
for 100 GeV jets. The $\kappa$ and $\lambda$ BDT and trainable $\kappa$ NN are also displayed.
The CNN is a two-input channel CNN with $\kappa=0.2$. The RNN is of type \ref{it4} using the $C/A$ distance. Both CNNs and the RNN noticeably outperform the $p_T$ weighted jet charge. The RecNN performs slightly worse than the RNN and CNNs, while the trainable $\kappa$ network only slightly outperforms jet charge. The $Q_{\kappa, \lambda}$ BDT outperforms jet charge and the trainable $\kappa$ NN but does not match the performance of the other NNs, particularly at low signal efficiency.
 \label{fig:best}}
\end{figure}

A comparison of the different RNN inputs is displayed in Fig. \ref{fig:RNNs}. All networks that take distance as input in Fig. \ref{fig:RNNs} use the $C/A$ distance. All configurations discussed above show slight improvement over jet charge. Configurations \ref{it1} and \ref{it5} perform only slightly better than jet charge, while the other RNNs perform noticeably better. We believe this is because configurations \ref{it2} through \ref{it4} all incorporate a measure of distance within the jet, similar to the CNN and RecNN displayed in Fig. \ref{fig:best}. 
  
 Performance of configuration \ref{it4} for the different distance measures is explored further in Fig.~\ref{fig:RNN5}. For configuration \ref{it2} the best performance was achieved for $N = 1$ with a subjet radius of 0.1.
 
\begin{figure}[t]
\centering
\includegraphics[width=80mm]{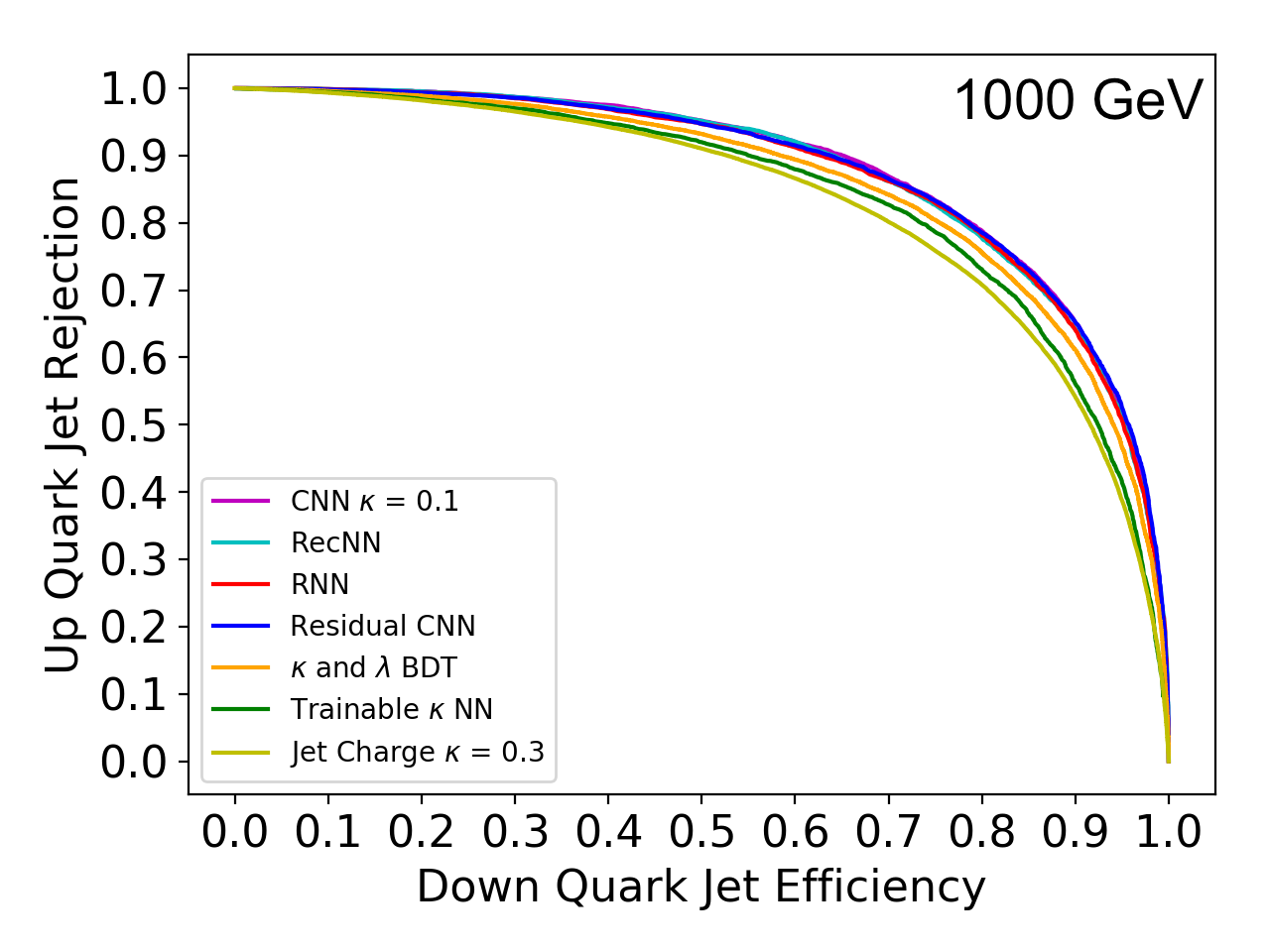}
\includegraphics[width=80mm]{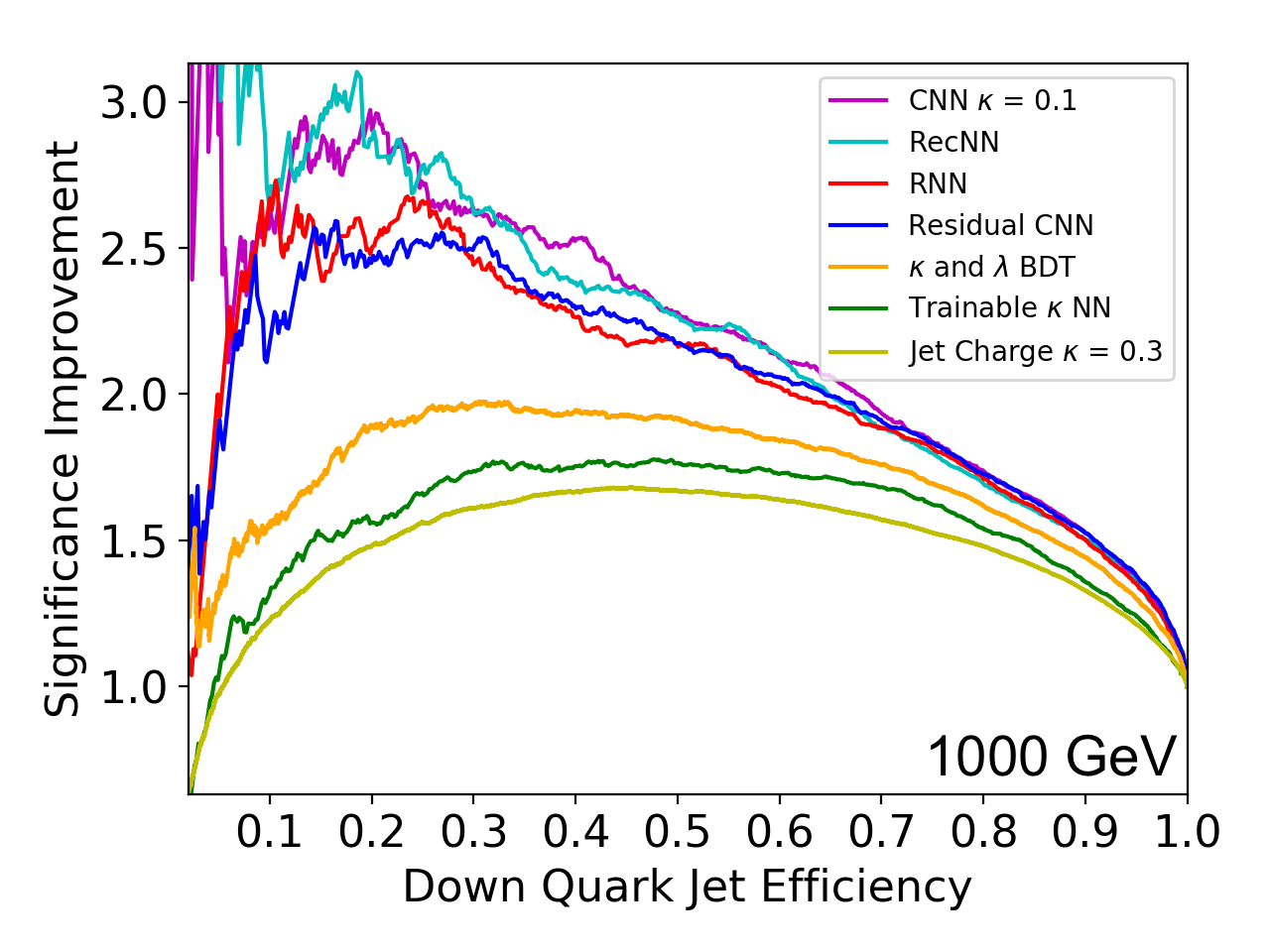}
\caption{Comparison of best performing recurrent (RNN), recursive (RecNN), and convolutional (CNN) neural networks with $p_T$-weighted jet charge at 1000 GeV. The $\kappa$ and $\lambda$ BDT and trainable $\kappa$ NN are also displayed. The improvement between the RNN, CNN or the RecNN and jet charge or the $Q_{\kappa, \lambda}$ BDT was larger than at 100 GeV.
\label{fig:best1000}}
\end{figure}  
 
We find that at low jet $p_T$ training is noticeably faster with the recurrent network than with jet images. At high jet $p_T$ the reverse is true.  Additionally, the training and performance of the recurrent network is sensitive to the ordering of the inputs and does not train unless they are sorted (for example, when inputs are ordered randomly the RNN is unable to distinguish the two samples). For the plots displayed in this paper, inputs are sorted in order of decreasing $p_T$, but we found that sorting by increasing distance from the jet axis is equally effective (which we expect since most jets have hardest particles toward the middle). We also found that including other extra information, in addition to the inputs of configuration \ref{it1}, inhibit training, sometimes to the point where the network is unable to reach an acceptance better than fifty percent. This suggests that including extra information in the RNN can actually hurt its performance. Various normalization configurations were tested, including zero centering and dividing by the standard deviation for a single jet, and zero centering and normalizing all channels across jets. Normalizing only the $p_T$ channel across jets was the only configuration that performed better than the raw \linebreak \clearpage \noindent vectors at 100 GeV. At 1000 GeV, this normalization was required to achieve an acceptance of better than fifty percent.

\subsection{Recursive Network Results}
The recursive network (RecNN) performed slightly worse than both the CNN and RNN for 100 GeV up versus down quark jets. Additionally, the embedding size required for effective training in this case was 25 parameters per particle, which is a larger embedding than the RNN. While our implementation of the RecNN was slower than the CNN or RNN, optimization measures such as dynamic batching implemented in \cite{Louppe:2017ipp} have been shown train faster than other implementations and make RecNNs feasible. However, the RecNN (like the CNN) can train with a small training set (16,000 events instead of 160,000), while the RNN does not achieve an acceptance of better than fifty percent for such a small training set.

 A comparison of the top performing convolutional, recurrent and recursive networks is shown in Fig. \ref{fig:best}. The area-under-the-ROC-curve (AUC) metric and the up quark efficiency at 50\% down quark efficiency are displayed in Table \ref{table:TABLE}.
 

\begin{table}[t]
\begin{center}
 \begin{tabular}{|c |c |c| c| c|} 
 \hline
 Network & \makecell{100 GeV \\ Up Quark Efficiency} & \makecell{100 GeV \\ AUC} & \makecell{1000 GeV \\ Up Quark Efficiency} & \makecell{1000 GeV \\ AUC} \\ [0.5ex] 
 \hline
 {\color{mplc}RecNN} & {\color{mplc} 0.085} & {\color{mplc}0.834}& {\color{mplc}0.049} & {\color{mplc}0.876} \\
 \hline
  {\color{mplm}CNN} & {\color{mplm}0.080} & {\color{mplm}0.837} & { \color{mplm}0.048} & { \color{mplm}0.879} \\
 \hline
 {\color{mplr}RNN} & { \color{mplr}0.079} & { \color{mplr}0.841} & {\color{mplr}0.054} & {\color{mplr}0.874} \\ 
 \hline
  {\color{mplb}Residual  CNN}  & { \color{mplb}0.078} & {\color{mplb}0.840} & {\color{mplb}0.053} & {\color{mplb}0.877} \\
 \hline
 {\color{mplo} $\kappa$ and $\lambda$ BDT}  & {\color{mplo}0.090} & {\color{mplo} 0.830} & {\color{mplo}0.068} & {\color{mplo}0.859} \\
 \hline
  {\color{mplg}Trainable $\kappa$ NN}  & {\color{mplg}0.104} & {\color{mplg} 0.815} & {\color{mplg}0.080} & {\color{mplg}0.841} \\
 \hline
   {\color{mply}Jet Charge}  & {\color{mply}0.109} & {\color{mply}0.810} & {\color{mply} 0.090} & {\color{mply}0.832} \\
 \hline
\end{tabular}
\caption{Up quark efficiency at 50\% down quark efficiency and area-under-the-ROC-curve (AUC) at 100 and 1000 GeV. Jet charge has $\kappa = 0.4$ at 100 GeV and $\kappa = 0.3$ at 1000 GeV. All NNs except the trainable $\kappa$ network noticeably outperform $pT$ weighted jet charge, as does the BDT. In the 100 GeV case, both CNNs and the RNN perform about equally well while the RecNN performs slightly worse. In the 1000 GeV case, the CNNs and RecNN give the best results, while the RNN performs slightly worse.}
 \label{table:TABLE}
 \end{center}
\end{table}

\subsection{Energy Dependence}
The results discussed above were all based on 100 GeV jets. The analysis was repeated for 1000 GeV jets. More precisely, up and down quark events were regenerated with $p_T$ between 1000 GeV and 1200 GeV, with all other parameters being the same.
We found discrimination power improves for all methods at higher $p_T$. This is of course expected and consistent with previous 
results~\cite{Krohn:2012fg,Waalewijn:2012sv,TheATLAScollaboration:2013sia}. Results are shown in Fig.~\ref{fig:best1000}. There was improvement in all methods, but the relative improvement of the RNN, CNN and RecNN over the $p_T$-weighted jet charge is larger at 1000 GeV than at 100 GeV. We also see that at 1000 GeV the RecNN and CNNs perform better than the RNN, in contrast to at 100 GeV where the RNN was best. Additionally, the improvement of the NNs over the $Q_{\kappa, \lambda}$ BDT is larger at 1000 GEV than at 100GEV.

\begin{figure*}[t]
\centering
\begin{subfigure}[t]{0.5\textwidth}
\centering
\includegraphics[width=\textwidth]{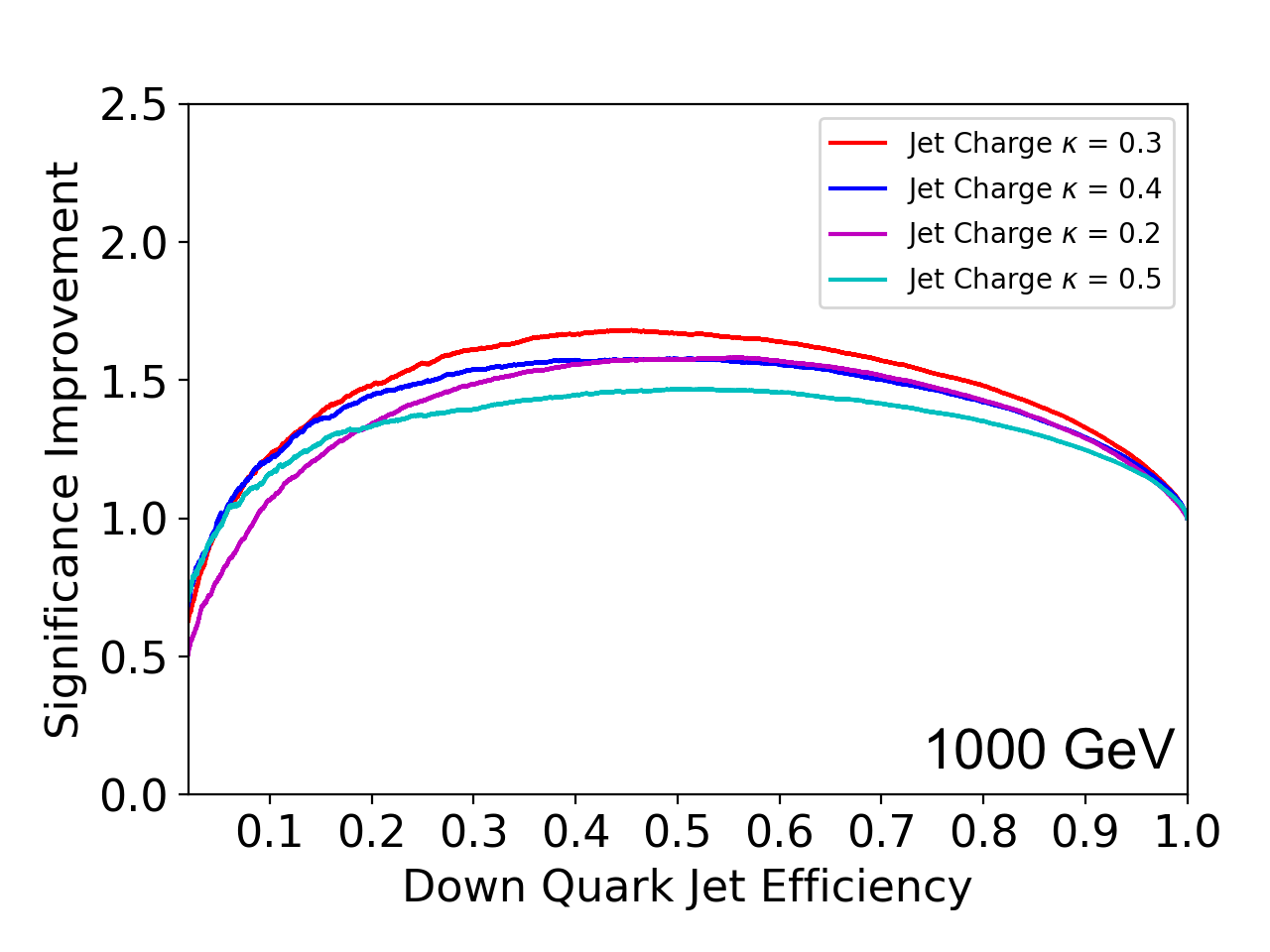}
\caption[]%
{{}}
\label{fig:1000GEVjetcharge}
\end{subfigure} 
\hfill
\begin{subfigure}[t]{0.49\textwidth}
\centering
\includegraphics[width=\textwidth]{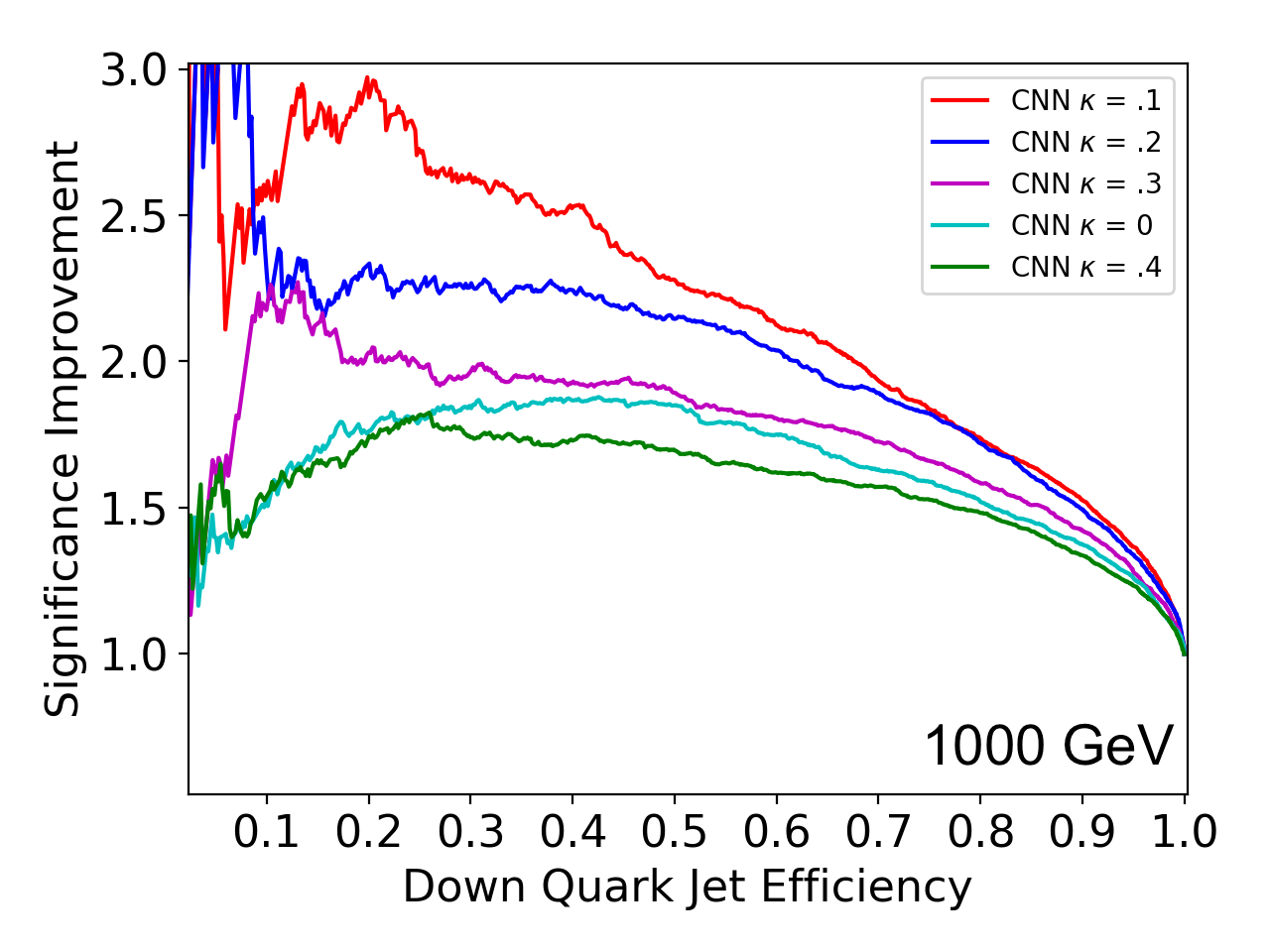}
\caption[]
{{}}
\label{fig:1000GEVCNNcharge}
\end{subfigure} 
\caption{(a) Comparison of jet charge for various $\kappa$ values at 1000 GeV. (b) Comparison of CNN performance for various $\kappa$ values at 1000 GeV.}
\end{figure*}

Figs.~\ref{fig:1000GEVjetcharge} and ~\ref{fig:1000GEVCNNcharge} try different values of $\kappa$ for the $p_T$ weighted jet charge and for the two-input-layer CNN. We see that the optimal $\kappa$ for both jet charge and the CNN decreases with energy. At 1000 GeV, the optimal $\kappa$ for the CNN is still smaller than the optimal $\kappa$ for jet charge.

\subsection{Quark/Gluon discrimination}
Finally, we examine how the network architectures that we have used for jet charge work for quark/gluon discrimination.
We compare our networks to each other as well as to the three-channel images used in \cite{Komiske:2016rsd} (which does not include jet charge), where one channel is total $p_T$, one is charged particle $p_T$ and the third
is particle multiplicity. For completeness, we also consider four-channel images with three channels as in \cite{Komiske:2016rsd} and 
a fourth having $p_T$-weighted jet change with $\kappa=0.2$ at 100 GeV and $\kappa=0.1$ at 1000 GeV (the same values as the best performing $\kappa$ in the up versus down quark case). We look at both 100 GeV and 1000 GeV jets.

\begin{figure*}[t]
\centering
\begin{subfigure}[t]{0.5\textwidth}
\centering
\includegraphics[width=\textwidth]{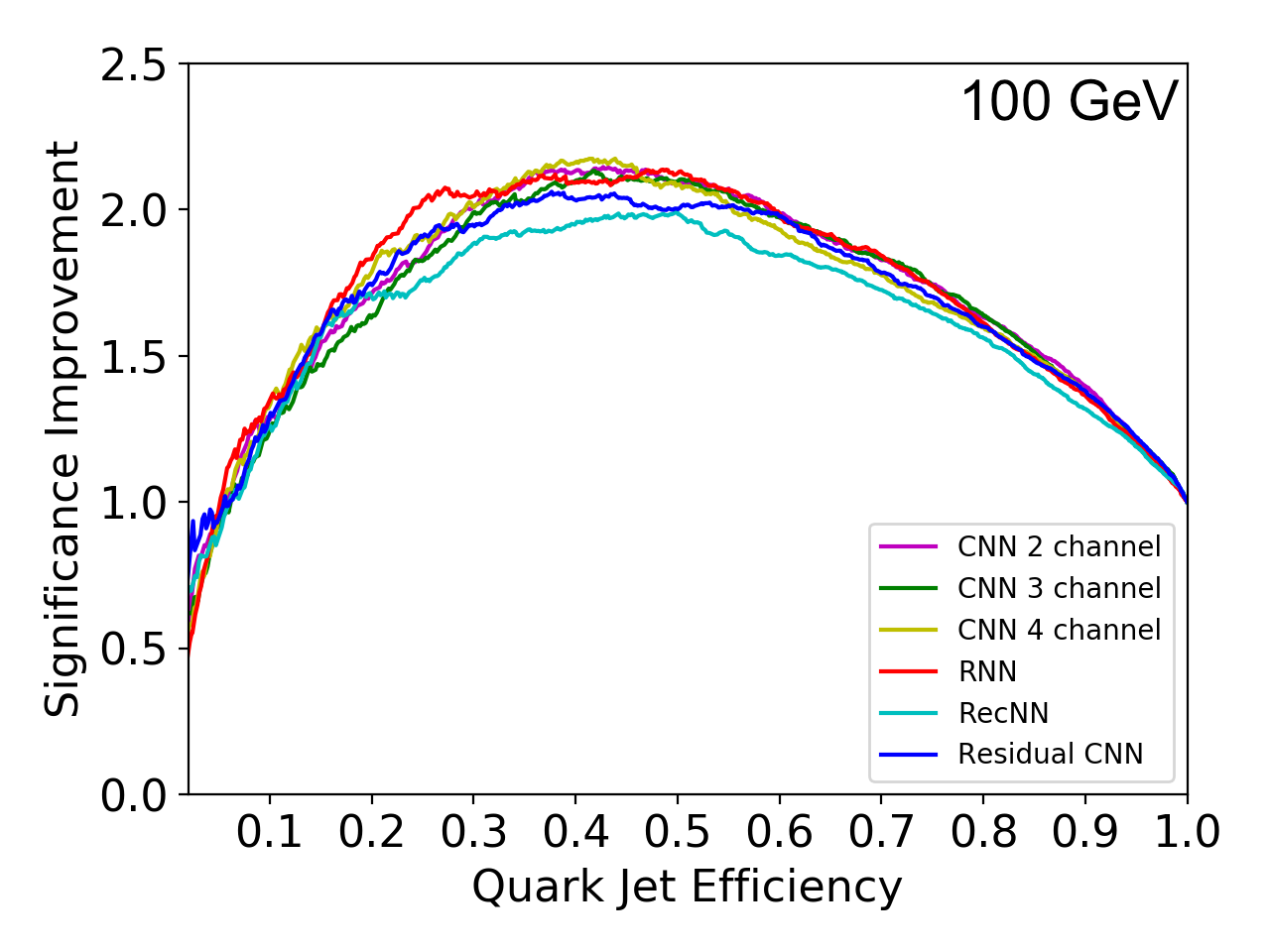}
\caption[]%
{{}}
\label{fig:QG100}
\end{subfigure} 
\hfill
\begin{subfigure}[t]{0.49\textwidth}
\centering
\includegraphics[width=\textwidth]{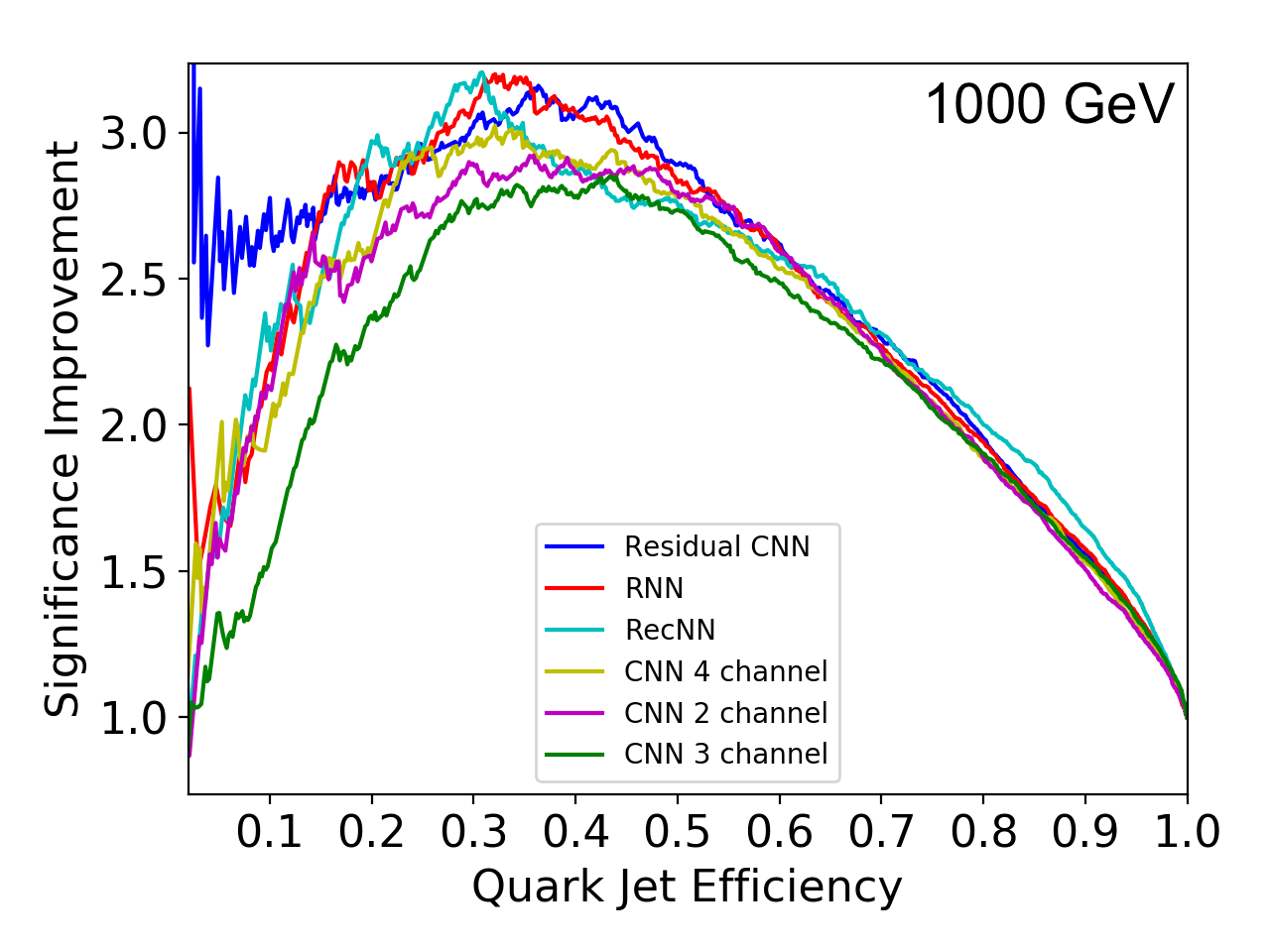}
\caption[]
{{}}
\label{fig:QG1000}
\end{subfigure} 
\caption{(a) Comparison of various network architectures for quark/gluon discrimination at 100 GeV. (b) Comparison of various network architectures for quark/gluon discrimination at 1000 GeV.}
\end{figure*}

Fig.~\ref{fig:QG100} is a comparison plot of the different methods. We see that most methods have comparable performance, with the exception of the recursive neural network that performs worse.
At 1000 GeV, the comparison is shown in Fig.~\ref{fig:QG1000}. We find in this case that the recurrent network does noticeably better than the jet images network. 
 

\section{Conclusions \label{sec:conc}}
In this paper we have applied techniques of modern machine learning to the problem of measuring the electric charge of a jet. In particular, we have used these networks to discriminate jets initiated by up-quarks (charge $Q=\frac{2}{3}$) from those initiated by down-quarks (charge $Q=-\frac{1}{3}$). The reference discriminator is the $p_T$-weighted jet charge~\cite{Krohn:2012fg} which has optimal performance for $\kappa \approx 0.4$ at 100 GeV and optimal performance for $\kappa \approx 0.3$ at 1000 GeV. The network architectures we considered include convolutional,  residual convolutional, recurrent and  recursive networks. We also studied boosted decision trees of $p_T$ and $\Delta R$ weighted jet electric charge. 

The CNNs are used to process jet images, with 2, 3 or 4 ``colors" (input channels) modeled
after the quark/gluon study in~\cite{Komiske:2016rsd}. We find these CNNs perform significantly better than the $p_T$-weighted jet charge. We also studied residual CNNs, which performed similarly to our other CNN (while there is some improvement in quark versus gluon discrimination at high energy and small signal efficiency, the improvement is not consistent across samples). 
For the recurrent networks we considered  a variety of different inputs. Recurrent networks take as input a list of variables associated with each particle, such as the 4-momenta or charge. We tried a number of different input sets and found that taking $\eta$, $\phi$, $p_T$, charge $Q$, and the $C/A$ clustering distance to the jet axis works the best. The performance of the recurrent network depends on its inputs: we find it is important to reduce the inputs from the raw 4-vectors to the energy and some distance measure. In principle, the network should learn this reduction, but doing so may require a very large network or enormously long training times. By processing the RNN inputs in this way, training is much faster and performance better. The improvement of RNNs with the inclusion of distance motivated a BDT study that used observables constructed from both $p_T$ and $\Delta R$ weighted electric charge as input, which improved performance over $p_T$ weighted jet charge alone. We also studied a recursive network with inputs ordered by clustering history, which performed similarly to the RNN and CNNs. Additionally, we tested a recursive network with multiple trainable $\kappa$'s, but this network barely outperformed jet charge. With the exception of this last network, all of our networks noticeably perform better than $p_T$-weighted jet change. While the BDT of $Q_{\kappa, \lambda}$ observables also outperforms jet charge, it does not match the performance of our neural networks, especially at high energy.

Our best networks can distinguish up and down quark jets significantly better than previous methods. At a 50\% down-quark efficiency working point, the networks allow us to reject all but 8\% of up-quark jets at 100 GeV (with a CNN or RNN) and all but 5\% of up-quark jets at 1000 GeV (with a CNN or a RecNN). This rejection rates improve on previous methods by almost a factor of 2 at high energy.  

Generally, discriminants that are useful for jet charge measurement are not infrared or collinear safe. For example, the $p_T$-weighted jet charge has this property, as do the multivariate methods  we use to study charge. These discriminants can still be measured, and some have been measured~\cite{TheATLAScollaboration:2013sia, Nachman:2014qma,TheATLAScollaboration:2015bgc, Aad:2015cua,CMS:2016yuu, Sirunyan:2017tyr,Tokar:2017syr}, with good agreement with theory. The importance of IRC safety in NN design and application is an interesting question that merits further investigation.

There are a few general lessons we have learned about networks from this study. At high signal efficiency, the neural networks that explicitly incorporate distance information (e.g. $\Delta R$ from the jet axis, pixel location in images, or distance from the jet's clustering history) perform about equally well. On the one hand, this may imply that there exist simple observables incorporating distance which perform as well as our neural networks. This motivated us to study some elementary attempts to include $\Delta R$ in observables, such as $Q_{\kappa, \lambda}$ (defined in eqn.~\ref{eqn:QLK}). Although $Q_{\kappa, \lambda}$ alone performs optimally for $\lambda = 0$ (which is just $p_T$-weighted jet charge), a BDT of such $Q_{\kappa, \lambda}$ observables with multiple values of $\lambda$ outperforms a BDT that only contains jet charge. This BDT study and the improved performance of the recurrent network when $C/A$ jet distance is explicitly included show that jet substructure can be more effectively used in jet electric charge classification. Specifically, distance information can be utilized to improve upon $p_T$-weighted jet charge in jet flavor classification, in both machine learning and more traditional observables. On the other hand, even our $Q_{\kappa, \lambda}$ BDT does not perform as well as the neural networks we studied, especially at higher $p_T$. This suggests that neural networks are able to fit a better function of distance than we can easily design, and/or that they are able to also utilize other information for performance gains. Therefore, we might conclude that searching for simple observables may not be worthwhile as the neural networks already perform well, have distance information, and can be used directly on data. 

 At low signal efficiency, which network performs best is dependent on what particle the jet is initiated by and the jet's energy. We found that with effective tuning of hyperparameters and normalization conventions all networks had similar performance. This suggests that while it is important to customize the size and parameters of a network to the specific application, in the case of up versus down jet identification neural networks that encode distances effectively should perform close to optimal. We see similar results in the quark gluon case. Since the networks perform equivalently, the difficulty of training the network should be an important consideration and should be customized to the particular application. An advantage of the CNN architecture is that it requires less modification with energy scale because the input representation size is fixed. An advantage of the RNN is that the input representation is smaller which can improve training time or memory usage, depending on implementation.

In conclusion, we have shown that machine learning can produce significant improvement in distinguishing up and down quark jets over traditional approaches. Our studies show that radial distance to the jet axis is one piece of information that can be utilized to contribute to this improvement. Our summary plots are in Figs.~\ref{fig:best} and \ref{fig:best1000}.
Neural networks that explicitly incorporate distance or clustering history are the most effective. Convolutional networks (like those used in \cite{Komiske:2016rsd}), recurrent neural networks, and recursive neural networks (like those used in \cite{Louppe:2017ipp}) perform very well.

\section*{Acknowledgements}
The authors would like to thank B. Nachman and J. Andreas for discussions, and the anonymous reviewer for suggesting the $Q_{\kappa, \lambda}$ study. MDS is supported in part by the Department of Energy under contract DE-SC0013607. Support for KF was provided in part by the Harvard Data Science Initiative and in part by the National Science Foundation Graduate Research Fellowship Program under Grant No. DGE1745303. The computations in this paper were run on the Odyssey cluster supported by the FAS Division of Science, Research Computing Group at Harvard University.

\bibliography{JetChargeML}

\bibliographystyle{utphys}

\end{document}